# THE FEASIBILITY OF ESTABLISHING A NUCLEAR PRESENCE
# AT
# WESTERN MICHIGAN UNIVERSITY

Prepared by

Andrew Thomas Goheen

Zayon Deshon Mobley-Wright

Rayne Symone Strother

Ryan Thomas Guthrie

December 6, 2023

Department of Business Communications

BCM 1420

Professor Tracey Corbett

Haworth College of Business, Western Michigan University

Kalamazoo, MI 49008

**TABLE OF CONTENTS**

**List of Figures**

**List of Tables**

## Executive Summary

As of this time, the State of Michigan utilizes a variety of different energy sources, including natural gas, coal, hydroelectric, nuclear, as well as wind and solar. This report aims to explore the utilization of nuclear energy on Western Michigan University's College Campus, exploring the feasibility of using this alternative source of energy to meet the campus community's long-term energy needs. Understanding that nuclear power is cost-competitive with other forms of electricity generation, the possibility of implementation is not fictitious. Numerous Universities have already implemented Small Modular Reactors (SMRs) to meet energy needs. Purdue University and Duke Energy are both uniquely qualified schools that have taken this "giant leap" toward a carbon-free energy future (Doty, 2022). As a school holding such a high regard for renewability and sustainability, Western Michigan University makes for a great candidate to experiment with nuclear energy.

## Intro to Nuclear Energy

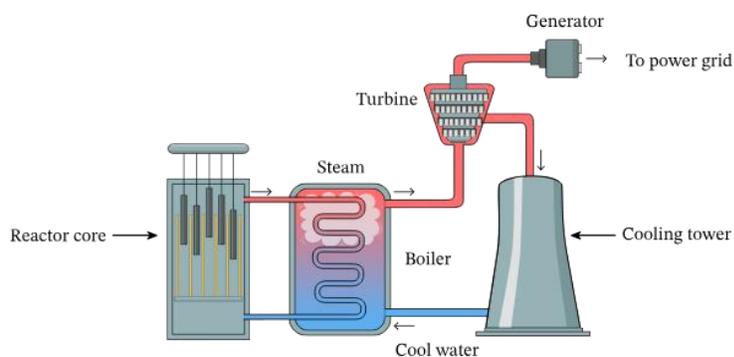

*Figure 1: Nuclear Energy Diagram*

Assuming protons and neutrons are commonly understood, they exist as the microscopic particles within atoms, when these particles are split, a remarkable amount of energy is released, this process is referred to as nuclear fission, occurring within what is known as nuclear reactors. The reaction within these atoms generates a surplus of heat, which is then used to create steam. The steam then spins large turbines, and in turn generates electricity. Nuclear energy produces an exponential amount of power without creating greenhouse gases or pollution as fossil fuels do. The United States has more than 100 reactors, although it creates most of its electricity from fossil fuels and hydroelectric energy. Nations such as Lithuania, France, and Slovakia create almost all their electricity from nuclear power plants. (Galindo, 2022)

**COLLECTED DATA**

**The Benefits of Nuclear Energy**

Nuclear energy provides benefits for the environment, the economy, and public safety. Environmental benefits can be seen through the removal of carbon and the recycling of nuclear waste. In terms of economic structure, benefits can be seen through an increase in job openings and through money being provided to both the government and local economies by the nuclear energy industry. Finally, the benefits of public safety can be seen through a decrease in deaths, as well as workplace accidents when it comes to fossil fuels and nuclear energy. In collecting the data, the following sources were used: An article by Martina Igini on the advantages and disadvantages of nuclear energy (Igini, 2023), and an article by the Office of Nuclear Energy on the advantages and challenges of nuclear energy (ONE, 2021).

<u>Environmental Benefits</u>

Nuclear energy has proven to be a clean and pollution-free power source with no greenhouse gas emissions. The energy source generates nearly 800 billion kilowatt hours of electricity per year while also only cutting down on 470 million metric tons of carbon each year (ONE, 2021). This is the equivalent of removing 100 million cars off the road. Many experts believe that nuclear energy is not only one of the cleanest sources of energy but also the power source responsible for the fastest decarbonization in history. This is seen through various countries showing a fast decline in carbon intensity after creating more sources of nuclear energy. Dealing with nuclear waste is also an issue that is far less impactful than that of climate change. This is because nuclear waste generated by nuclear energy production can be recycled. The recycled fuel can then be treated and put into another reactor (Igini, 2023)

| GREENHOUSE GAS EMISSIONS (TONS) | | | | | | | |
| --- | --- | --- | --- | --- | --- | --- | --- |
| Measured in emissions of CO2-equivalents per gigawatt hour of electricity over the lifecycle of 1 gigawatt-hour is the annual electricity consumption of 160 people in the EU | | | | | | | |
| **Coal** | **Oil** | **Natural Gas** | **Biomass** | **Hydropower** | **Nuclear Energy** | **Wind** | **Solar** |
| **820** | **720** | **490** | **230** | **34** | **3** | **4** | **5** |

Table 1: Greenhouse Gas Emissions Per Energy Source

<u>Economics and Public Safety</u>

Concerning employment, the nuclear energy industry opens almost half a million jobs in the U.S. alone. The industry also provides around $60 billion to the U.S. in gross domestic products annually, as well as billions of dollars to local economies through tax revenues (ONE, 2021). The plants that open can hire up to 700 people and give the workers a salary that is 30% higher than the local average. Not only does the nuclear energy industry support the economy but the safety of people as well. While fossil fuels are responsible for 1 in 5 deaths worldwide and for 8.7 million deaths in 2018 alone, in the past 70 years since nuclear energy began being used, only

three incidents have threatened public safety. Only one of these three incidents caused any deaths directly, which was Chornobyl (Igini, 2023).

| DEATH RATE FROM ACCIDENTS AND AIR POLLUTION Measured as deaths per terawatt-hour of energy production. 1 terawatt-hour is the annual energy consumption of 27,000 people in the EU | | | | | | | |
|---|---|---|---|---|---|---|---|
| **Coal** | **Oil** | **Natural Gas** | **Biomass** | **Hydropower** | **Nuclear Energy** | **Wind** | **Solar** |
| 24.6 | 18.4 | 2.8 | 4.6 | 0.02 | 0.07 | 0.04 | 0.02 |

Table 2: Death Rate Per Energy Source

## Negative Effects of Fossil Fuels

Fossil fuels cause many issues, not only within the environment but economically as well. These issues can be seen in the environment with ocean acidification, severe weather, and the rise in sea level. It can also be seen in pollution through air pollution, water pollution, and plastic pollution. In collecting the data, the following sources were used: An article by Savannah Bertrand on the climate, environment, and health impacts of fossil fuels (Bertrand, 2021), and an article by Hannah Ritchie and Pablo Rosado on natural disasters (Ritchie & Rosado, 2022)

### Environment

Our oceans are very fragile, almost a quarter of the carbon emissions caused by the burning of fossil fuels are absorbed into the ocean. This absorption causes a change to the ocean's pH levels, leaving the water with more acidity. The increase in acidity has been going on for over a century and has increased by 30%, posing a threat to marine life, coral reefs, and the economy. Severe weather events are also an issue created by the burning of fossil fuels. The impact of climate change created by fossil fuels is what causes these events. These weather events can lead to billions of dollars being used per event to recover. From 2016 to 2020 in the U.S. alone, the cost was over $606 billion (Bertrand, 2021). Not only are weather events a major cost, but they also can lead to many deaths. Globally there are 45,000 deaths per year due to severe weather (Ritchie & Rosado, 2022). Climate change is also affecting the sea level, causing it to rise as glaciers and land-based ice sheets melt. Since the 1800s, the sea level has risen 9 inches, which has caused more frequent flooding, storm surges, and saltwater intrusion. The rise in sea level can also lead to increased costs for coastal communities. 40% of the U.S. population lives along the coast, and the defense against the rise of sea level can cost these communities over $400 billion over the next 20 years (Bertrand, 2021).

<u>Pollution</u>

Clean air is a necessary resource, but unfortunately, the burning of fossil fuels creates hazardous air pollutants. These pollutants include sulfur dioxide, nitrogen oxides, particulate matter, carbon monoxide, and mercury. Not only are these chemicals hazardous to the environment, but the pollutants also are harmful to human health. When it comes to the environment, the burning of fossil fuels can cause acid rain, eutrophication, damage to crops and forests, and harm to wildlife. When it comes to human health, air pollution can cause asthma, cancer, heart disease, and premature death. The use of fossil fuels also causes water pollution through both fracking and oil spills. With fracking, each fracking well uses between 1.5 million to 16 million gallons of water. The water used during the fracking process can be toxic, containing substances such as arsenic, lead, chlorine, and mercury. This toxic water can contaminate both groundwater and drinking water. Plastic pollution is also a large concern regarding the use of fossil fuels. The concerns include the dumping of plastic waste into the oceans and the effect on the climate that it holds. Over 99% of plastic is made from fossil fuels, with 300 million tons of plastic waste being produced globally and annually. 14 million tons of this waste is dumped into the oceans. The dumping of this waste kills wildlife and pollutes the environment. When it comes to the effect on the environment it holds, the U.S. plastic industry itself produces 232 million tons of carbon dioxide annually. This production of carbon monoxide is expected to grow exponentially by 2030 (Bertrand, 2021).

**Implementation**

In recent years, universities have become increasingly interested in sustainability and exploring eco-friendly ways to meet their energy requirements. One such solution that has been gaining attention is the use of nuclear energy. The Delft University of Technology has already implemented nuclear energy for both power generation and district heating, setting an example for others to

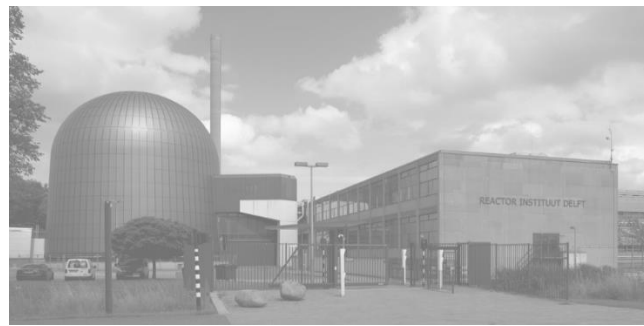
*Figure 2: Delft University Nuclear Facility*

follow. However, introducing nuclear power at a university like Western Michigan would require strict adherence to regulatory compliance and safety protocols, as well as active engagement with the community. To determine the most suitable nuclear energy implementation strategy, Western University should engage in comprehensive strategic planning involving all stakeholders. While students may not have a direct role in implementing nuclear projects, their active participation and advocacy can positively contribute to a well-informed and responsible approach towards such initiatives. Collaboration with experts, adherence to regulations, and ongoing communication with stakeholders are crucial for successfully and responsibly integrating nuclear energy into a university's energy portfolio.

**Comprehensive Analysis**

Based on the information provided, it should be assumed that the integration of nuclear energy within college campuses has garnered considerable attention as a viable means of sustainable power generation. By exploring the technical, environmental, social, and economic aspects, this analysis aims to provide a comprehensive understanding of the potential impact and suitability of nuclear energy adoption within an educational institution.

Implementing nuclear energy on a college campus involves various technical considerations. The installation of a nuclear reactor, such as a small modular reactor (SMR), demands rigorous safety measures, licensing procedures, and specialized infrastructure. The technical team's expertise, maintenance protocols, and emergency response plans are pivotal in ensuring safe operations. Additionally, the campus's existing power grid and infrastructure must be evaluated for compatibility and capacity to integrate nuclear energy seamlessly. Often regarded as a low-carbon alternative, nuclear energy presents a cleaner energy source compared to fossil fuels. However, concerns persist regarding radioactive waste disposal and the environmental impact of uranium mining. Comprehensive waste management strategies and adherence to stringent safety protocols are imperative to mitigate environmental risks. Evaluating the life-cycle emissions, including construction, operation, and decommissioning, provides a holistic understanding of nuclear energy's environmental footprint.

**Survey**

In our survey conducted among students, instructors, and staff members at Western Michigan University, diverse responses were gathered concerning government funding for research and development in nuclear energy. When faced with the question, **"Would you support increased government funding for research and development in nuclear energy?"** 50% of participants expressed strong support, 35% were moderately supportive, 7% were neutral, and 8% indicated no support for nuclear funding. We identified that a lack of information regarding nuclear energy influenced these results. Based on past survey analysis information, individuals with a better understanding of nuclear energy were more likely to express support.

Continuing our survey following a nuclear energy presentation within the Bronco community, we introduced the question, **"In the future, do you see yourself being more or less supportive of nuclear energy?"** The results revealed a curve, with 70% of the community expressing a level of support for nuclear energy. Interestingly, the remaining 30% of respondents who identified as non-supporters voiced significant concerns. Their hesitations were tied to the global impact nuclear energy could have at Western Michigan University, worries related to waste management, safety issues, and potential funding challenges.

# Budget

| Net Position as of June 30 (in millions) | 2023 | 2022 Restated | 2021 |
|---|---|---|---|
| Net investment in capital assets | $ 527.0 | $ 501.8 | $ 497.5 |
| Restricted | 5.5 | 6.5 | 6.5 |
| Unrestricted - operations | 367.6 | 341.6 | 332.2 |
| Unrestricted - pension activity | (65.0) | (138.6) | (148.9) |
| Unrestricted - other postemployment benefit (OPEB) activity | (201.5) | (241.1) | (271.1) |
| Total net position | $ 633.6 | $ 470.2 | $ 416.2 |

| Net Position Excluding Pension and OPEB Accounting as of June 30 (in millions) | 2023 | 2022 Restated | 2021 |
|---|---|---|---|
| Net investment in capital assets | $ 527.0 | $ 501.8 | $ 497.5 |
| Restricted | 5.5 | 6.5 | 6.5 |
| Unrestricted - operations | 367.6 | 341.6 | 332.2 |
| Total net position excluding pension and OPEB | $ 900.1 | $ 849.9 | $ 836.2 |

Table 3: Western Michigan University Audited Financial Report - June 2023

**Year 1: Getting Started**

- Necessary licenses: $5 million

- Research: $5 million

- Legal and consulting fees: $2 million

**Year 2-3: Building the Infrastructure**

- Building the nuclear reactor facilities: $150 million

- Connecting the plant to the electricity grid: $20 million

- Setting up cooling systems and managing water: $15 million

- Implementing security measures: $10 million

**Year 4: Buying Technology and Equipment**

- Buying nuclear reactor technology: $50 million

- Setting up control systems: $7 million

- Turbines and generators: $20 million

**Year 5: Training the Workforce**

- Training programs for operators and technicians: $5 million

- Initiatives to develop the workforce: $3 million

**Year 6-7: Starting Operations**

- Initial costs for operations and maintenance: $25 million

- Buying fuel: $15 million

**Year 8: Engaging with the Community & Western Michigan University**

- Outreach engagement to the community: $2 million

- Public awareness: $2 million

**Year 9: Research and Innovation**

- Research and development for further nuclear technologies and projects: $5 million

**Year 10: Final Adjustments**

- Incident Fund: $20 million

- Final adjustments/ unexpected expenses: $10 million

The projected cost for the entire 10-year span is $327 million, and our budget is structured with a yearly spending average of $39 million. This 10-year breakdown has been carefully designed to afford us the necessary time for adjustments that may arise during the project. Each annual deployment plan has been set to enable a focused and detailed approach, ensuring that every aspect is constructed with the best quality. Given the complexity of completing this project, each year needs careful attention to detail to ensure it's done successfully.

*The following numbers are subjective and by no means reflect accurate spending by the University.*

## Conclusions

Nuclear energy, which is more than capable of producing the electricity that powers your devices and illuminates your surroundings, is frequently undervalued in the United States. This undervaluation is mainly attributed to a lack of awareness regarding its true capabilities. The full potential of nuclear energy remains obscured due to a general lack of knowledge and awareness. With a better understanding of the great benefits that come with nuclear energy, it can be assumed that most people would be highly inclined to consider nuclear energy as a viable and sustainable option. Transitioning to more nuclear energy would bring about substantial benefits swiftly, particularly when considering the environmental footprint.

## Recommendations

Based on the report's findings, it is recommended that colleges explore the potential of implementing nuclear power to meet their energy needs. Universities with nuclear reactors can benefit from a consistent and reliable power supply, which is advantageous for research activities requiring stable and continuous power. This contributes to energy autonomy for the university and guarantees that power failures do not disrupt vital experiments and research. Western Michigan University can consider collaborating with energy companies to investigate the use of Small Modular Reactors. The Cook Nuclear Plant, located on the coast of Lake Michigan, is a nearby facility with two Westinghouse pressurized water reactors licensed until 2034 and 2037, which could be a viable option. (Westinghouse, 2023)